\documentclass[12pt]{article}

\usepackage{amssymb,amsmath}
\usepackage{graphbox}
\usepackage{graphicx}
\usepackage{slashed}

\begin{document}

\begin{titlepage}

\begin{flushright}
arXiv:2609.13686
\end{flushright}
\vskip 2.5cm

\begin{center}
{\Large \bf Lorentz Violation and Vector Meson Dominance}
\end{center}

\vspace{1ex}

\begin{center}
{\large Brett Altschul\footnote{{\tt altschul@mailbox.sc.edu}}}

\vspace{5mm}
{\sl Department of Physics and Astronomy} \\
{\sl University of South Carolina} \\
{\sl Columbia, SC 29208}
\end{center}

\vspace{2.5ex}

\medskip

\centerline {\bf Abstract}

\bigskip

Vector meson dominance models describe the mixing of photons with vector mesons; the hadronic admixture
frequently dominates the interactions of photons with nucleons. This mixing also makes it possible to
test for new physics affecting the vector mesons using photon measurements.
The possibilities for new physics to be constrained in this indirect fashion include
potential Lorentz and CPT violations in the hadronic effective action. The characteristic scale of
bounds on Lorentz violation coefficients in the $\rho^{0}$ meson sector is two orders of magnitude weaker than
the corresponding photon bounds, three orders of magnitude weaker for $\phi$ vector mesons, and
four orders of magnitude weaker for $\omega$ mesons.

\bigskip

\end{titlepage}

\newpage

\section{Introduction}

More than 120 years have elapsed since the first enunciation of the special theory of relativity.
Throughout all that time, there has, quite naturally, been an enduring interest in the question of
whether the Lorentz symmetry underlying relativity is an exact property of the theory, or whether
it is merely an extremely accurate approximation. This question has even arguably become more
interesting over time, since it is now well understood that approximate symmetries---such as
isospin, parity, and time reversal---can be exceedingly important. Understanding the mechanisms by
which such symmetries are feebly broken has taught us a tremendous amount about the fundamental laws
of physics.
Questions about the possible mechanism of Lorentz symmetry breaking have evolved a
great deal over the last three decades, thanks to the maturation of effective field theory (EFT)
techniques. Quantitative investigations into how well Lorentz invariance has been confirmed have
exploded, both in number and in quality, since it has been realized how rich the landscape of possible forms
for Lorentz violation actually is.

The fundamental laws of physics, as we presently understand them, are formulated using two quite different
frameworks. The standard model, which describes elementary particle physics, is a relativistic quantum
field theory. In contrast, general relativity is a description of gravitation in terms of spacetime
geometry. The most basic and challenging problem in fundamental physics is to integrate these two
formalisms to give a quantum theory of gravity. Nevertheless, in spite of the profound differences between
how we understand particle physics and gravitation, the two basic theories actually have a number of key
features in common. These features include a number of symmetries that, as yet, appear to
be exact; both the standard model and general relativity are invariant under spatial rotations, Lorentz boosts,
and the combined discrete symmetry of charge conjugation, parity, and time reversal.
Experiments have not produced any compelling evidence challenging this paradigm of $SO(3,1)$ Lorentz
symmetry; however, if any deviations from Lorentz invariance were to be conclusively discovered,
that would obviously be of the deepest significance, and it would provide a completely new avenue for
studying and understanding the fundamental laws of the Universe.

With the modern machinery of EFT, it is possible (if not always straightforward) to describe
putative deviations from these common symmetries, in both particle and gravitational physics.
The most general EFT used for describing Lorentz-symmetry-breaking effects
is known as the standard model extension (SME)~\cite{ref-kost1,ref-kost2}. Because of the connections
between Lorentz and CPT symmetries~\cite{ref-greenberg}, the SME is also the general EFT for describing
CPT violation in quantum field theories that are well defined (in the sense of having a unitary $S$-matrix).
The full SME, in its greatest generality, contains infinite
towers of Lorentz-violating field operators. However, the Lagrange density for a restricted theory, called the
minimal SME, contains only operators constructed out of known fields that are gauge invariant,
energy-momentum conserving, and (for the particle physics sector of the theory) of mass dimension four or less,
making them
renormalizable~\cite{ref-kost3,ref-berr,ref-collad-3,ref-collad-2,ref-collad-1,ref-gomes,ref-anber,
ref-ferrero3,ref-altschul43,ref-altschul44,ref-felipe,ref-ageev}.
Minimal SME operators actually closely resemble the ones that appear in the usual actions for the standard
model and general relativity. The key difference is, while Lorentz-invariant operators have no free Lorentz
indices, Lorentz-violating operators may have indices that are not contracted internally.
The Lagrange density for the theory may be written so that the coefficients of the various Lorentz components
of a field operator come together to make vector- or tensor-valued objects, which represent preferred
directional backgrounds
in spacetime. If Lorentz symmetry is broken spontaneously, then these background
structures are proportional to the vacuum expectation values of the vector- or tensor-valued dynamical fields
responsible for the breaking.
The finite number of operators that make up the minimal minimal SME are a very useful framework for comparing the
results of experimental Lorentz tests performed in different ways and using different types of quanta.
In fact, experiments in many
different physical regimes have been used to place bounds on the coupling constants of the SME, and
information about the current experimental constraints on the parameters of the SME
are collected in~\cite{ref-tables}.

EFT is, of course, also very important to our understanding of other kinds of physical
phenomena. This is particularly true in hadronic physics. The fundamental physical theory of the strong
interactions---quantum chromodynamics (QCD)---is well established, but it can be challenging to relate
the QCD action, which is formulated in terms of quarks and gluons, to low-energy physical observations.
There have been some analyses of quark-level SME coefficients directly, in terms of
partons~\cite{ref-lunghi4,ref-lunghi3,ref-lunghi5,ref-lunghi2},
leading to direct particle-accelerator bounds in the top quark sector of the SME~\cite{ref-carle,ref-belyaev},
but these techniques are necessarily rather crude compared with other methods even in the top
sector~\cite{ref-altschul14,ref-altschul45}.
Because the QCD coupling is large at small momenta and the observable quanta
of the low-energy theory are composite mesons and baryons, a variety of EFTs have been developed to address
different hadronic regimes and phenomena. It is a worthwhile question to examine how the SME formalism
interacts with these other EFTs.

The simplest extension of the SME apparatus to an EFT context, rather than working directly with the
fundamental standard model fields,
is essentially as old as the SME formalism itself. Many of the most precise tests of
isotropy in the early years of this century came from atomic clock experiments measuring nuclear transitions.
The extremely tight bounds resulting from experiments like these are obviously not straightforward to
relate to the underlying quark and gluon parameters of the SME; so instead it was natural to express
them in terms of effective Lorentz violation coefficients defined for the composite nucleon
fields~\cite{ref-kost6}. In fact, even relating the results of atomic hyperfine and Zeeman splitting
measurements to individual nucleon coefficients requires an effective model of how the magnetic moment
of the nucleus is carried by individual nucleons, although a simple modification of Schmidt
model~\cite{ref-margenau} generally suffices for even-odd nuclei. Moreover, for very simple nuclei, it
may be possible to translate back and forth between effective SME coefficients for the constituent
nuclei and the whole nucleus, but so far work in this area has been very limited~\cite{ref-altschul38}.
Some similar, albeit incomplete, analyses have started to extend the hadron-level EFT approach to Lorentz and
CPT violation in the $\Delta$-resonance sector~\cite{ref-gomes3,ref-gomes4,ref-gomes2}.

More elaborate EFTs are designed to capture key features of the strong interaction and enable, within
their appropriate regimes, quantitative calculations with hadrons. One of the most
important hadronic EFTs is chiral perturbation theory ($\chi$PT).
As a low-energy theory, $\chi$PT is structured around the
properties of the pseudo-Goldstone bosons of chiral symmetry breaking---the isovector pions (and, when
the strange quark is included, the kaons as well). This theory provides the best current prospects for
understanding Lorentz violation for these light mesons, and also for translating the extremely tight
bounds on the effective SME coefficients for baryons and nuclei into contraints on the
parameters for their color-confined quark and gluon
constituents~\cite{ref-noordmans2,ref-kamand1,ref-noordmans1,ref-kamand2,ref-altschul46}.

Just as $\chi$PT is centered around the physical phenomenon of spontaneous chiral symmetry breaking
(one of the most important features of QCD), vector meson dominance (VMD)
models originated out of the physical observation that the interactions between hard photons
and nucleons could be much stronger than the naively
calculated electromagnetic form factors of the nucleons would imply.
Instead, there were interaction strengths more characteristic of strong interactions. VMD models
explain this as a result of the photon mixing with vector mesons, so that it is the meson component
of a dressed photon state that is principally responsible for the interactions of energetic photons with
other hadrons. The mesons that the photon mixes with must have the same $J^{PC}=1^{--}$ as the photon
itself. This makes the most important contributors to VMD the $\rho^{0}$, $\omega$, and $\phi$ mesons,
with masses of approximately 775, 782, and 1020 MeV, respectively. Details of the history,
experimental basis, and formalism of VMD were laid out nearly half a century ago in
Ref.~\cite{ref-bauer}; see also Ref.~\cite{ref-oconnell}.

In this paper, we shall explore a number of different ways in which the basic EFT for describing
VMD may be modified in accordance with the SME paradigm of writing down all permitted operators,
unconstrained by Lorentz or CPT symmetry, that may be built out of the theory's known quantum fields.
Our analysis is organized as follows. In section~\ref{sec-Lag}, the Lagrange densities for the electromagnetic
sector, meson sector, and their mixing are laid out---including Lorentz-violating terms not present in the
usual description of VMD. In section~\ref{sec-exp}, we look at the experimental bounds on these coefficients.
Section~\ref{sec-concl} presents a summary of the
paper's conclusions and outlook for further progress and improved bounds in the future.

\section{Effective Lagrange Densities}
\label{sec-Lag}

The Lagrange density for the free photon sector of the minimal SME is
\begin{equation}
\mathcal{L}_{A}= -\frac{1}{4}F^{\mu\nu}F_{\mu\nu}
-\frac{1}{4}k_{F}^{\mu\nu\sigma\tau} F_{\mu\nu}F_{\sigma\tau}
+\frac{1}{2}k_{AF}^{\mu}\epsilon_{\mu\nu\sigma\tau}F^{\nu\sigma}A^{\tau}.
\end{equation}
The four $k_{AF}$ terms are odd under CPT, and they are all extremely tightly bounded by measurements of the
birefringence from cosmological photon sources. The $k_{F}$ terms---of which there are nineteen, since
the four-index $k_{F}$ tensor has the symmetries of the Riemann curvature tensor and a vanishing double
trace---are CPT even. Of these, ten components (those corresponding in structure to a Weyl curvature
tensor) are also very tightly bounded by cosmological birefringence observations. The remaining nine (with
the symmetries of a Ricci tensor) represent a skewing of the ``natural'' Cartesian coordinates of the
electromagnetic sector relative to other sectors of the theory. They are also all well constrained, but
not so tightly as the terms that generate birefringence.

Many of the strongest bounds on the coefficients of the SME come from astronomical observations.
There are two things available astrophysically that are not possible in terrestrial laboratories: very
high energies and very long propagation distances. It is the latter that makes the tightest bounds in the
photon sector of the minimal SME possible; polarimetric measurements of radiation coming from cosmologically
distant sources take advantage of the fact that over megaparsec and gigaparsec distances even a tiny difference
in phase speed between two propagation eigenstates can lead to a large change in the polarization of a
wave.

It is by no means a novel observation that there are additional possibilities for Lorentz violation when
dealing with a massive vector field. There are Lorentz-violating operators of mass dimension two---generalizations
of the ordinary Proca mass term $\frac{1}{2}m_{V}^{2}V^{\mu}V_{\mu}$. However, previous explorations of these
possible terms in the context of a possible very small photon
mass~\cite{ref-gabadadze,ref-dvali,ref-altschul8}
have had a different character from what
will be relevant in a discussion of vector mesons for which the Lorentz-invariant masses are quite sizeable.
With a purported Lorentz-violating photon mass of, say, the form $-\frac{1}{2}m_{A}^{2}A_{j}A_{j}$, the smallness
of this term is really related only to the smallness of the mass scale involved. For mass-dependent
phenomena, the physical Lorentz violation would actually be quite large, since the relative sizes of the mass
parameters associated with the several components of the field are very different; with the example choice
of mass term, the mass parameter for each spacelike component of the photon field is $m_{A}$, but it
vanishes for the timelike component $A^{0}$. However, this not a problem for phenomenology, because the
Lorentz-violating mass term is still small in an absolute sense.
In contrast, for a vector meson like the $\rho^{0}$, with a substantial mass of around 775 MeV, Lorentz-violating
differences between the mass parameters for the different polarization states must be tiny fractions of the
average mass to maintain consistency with the known approximate validity of Lorentz symmetry.

With these considerations in mind, we have the a Lagrange density, without interactions, for each of the
relevant vector meson fields $V$,
\begin{equation}
\mathcal{L}_{V}= -\frac{1}{4}\left(G^{V}\right)^{\mu\nu}G^{V}_{\mu\nu}
-\frac{1}{4}k_{G^{V}}^{\mu\nu\sigma\tau} G^{V}_{\mu\nu}G^{V}_{\sigma\tau}
+\frac{1}{2}k_{VG}^{\mu}\epsilon_{\mu\nu\sigma\tau}\left(G^{V}\right)^{\nu\sigma}V^{\tau}
+\frac{1}{2}M^{V}_{\mu\nu}V^{\mu}V^{\nu}.
\end{equation}
where the meson field strength is $G^{V}_{\mu\nu}=\partial_{\mu}V_{\nu}-\partial_{\nu}V_{\mu}$,
and the possibly Lorentz-violating mass matrix is
\begin{equation}
M^{V}_{\mu\nu}=m_{V}^{2}g_{\mu\nu}+\mathfrak{m}^{V}_{\mu\nu}.
\end{equation}
The $k_{G^{V}}$ and $k_{VG}$ tensors parameterize types of Lorentz violation in the meson sector, and they have
the same structure as their electromagnetic analogues. The $\mathfrak{m}^{V}$ tensor is traceless (the trace
term of $M^{V}$ being absorbed into the standard Proca mass sqaured $m_{V}^{2}$), and
it represents the mass-like Lorentz-violating interaction for the meson species.
Lorentz violation for massive vector fields have also been studied using the Stueckelberg
formalism~\cite{ref-ferreira2,ref-cambiaso}.

One might hope, in light of the photon-meson mixing outlined below,
that measurements of photons would provide an avenue for exploring
the possibility of Lorentz-violating mass terms for the vector mesons with which the photons mix. However,
for real photons this turns out to be more difficult than it might initially
appear, as a consequence of gauge invariance. In the basic model, there is no way to
write a gauge-invariant operator that involves the two-index symmetric tensor $M^{V}$ and the dressed photon
field operators that has the correct dimensions to appear in the effective action. Structurally, the only
bilinear photon term would have to be proportional to $M^{V}_{\mu\nu}M^{V}_{\sigma\tau}F^{\mu\sigma}F^{\nu\tau}$.
But this operator, as written, has engineering dimension eight, and there is no mass scale in the simplest VMD
theory to divide by and thus correct this to produce a dimension-four operator such as could appear in the Lagrange
density.

Nevertheless, since the simple VMD theory is only an approximation---albeit often a very good one for understanding
photon-hadron interactions---it is worth examining the structure of this term a bit further.
Writing again the mass-squared matrix $M^{V}_{\mu\nu}$ as $m_{V}^{2}g_{\mu\nu}+\mathfrak{m}^{V}_{\mu\nu}$ and
expanding to leading order in the Lorentz violation, we consider an effective $k_{F}$-type term
\begin{eqnarray}
\mathcal{L}_{\mathrm{mass}} & = & -\frac{1}{4\Lambda_{V}^{4}}\left(m_{V}^{2}g_{\mu\nu}+\mathfrak{m}^{V}_{\mu\nu}\right)
\left(m_{V}^{2}g_{\sigma\tau}+\mathfrak{m}^{V}_{\sigma\tau}\right)F^{\mu\sigma}F^{\nu\tau}
\label{eq-Lmass} \\
& \approx & -\frac{m_{V}^{4}}{4\Lambda_{V}^{4}}F^{\mu\nu}F_{\mu\nu}-\frac{m_{V}^{2}}{2\Lambda_{V}^{4}}
\mathfrak{m}^{V}_{\mu\nu}g_{\sigma\tau}F^{\mu\sigma}F^{\nu\tau},
\end{eqnarray}
for some scale $\Lambda_{V}\gg m_{V}\sim 1$ GeV that characterizes the breakdown of the VDM model.
Formally, this is another layer of EFT, with the introduction of a new cutoff scale $\Lambda_{V}$ for each
species.
The key feature of the Lagrange density $\mathcal{L}_{\mathrm{mass}}$ is the presence of an effective $k_{F}$
coefficient
\begin{equation}
\label{eq-kmF}
k_{\mathfrak{m}F}^{\mu\nu\sigma\tau}=\frac{m_{V}^{2}}{2\Lambda_{V}^{4}}\left[
\left(\mathfrak{m}^{V}\right)^{\mu\sigma}g^{\nu\tau}-\left(\mathfrak{m}^{V}\right)^{\nu\sigma}g^{\mu\tau}
+\left(\mathfrak{m}^{V}\right)^{\nu\tau}g^{\mu\sigma}-\left(\mathfrak{m}^{V}\right)^{\mu\tau}g^{\nu\sigma}\right].
\end{equation}
The form of the effective $k_{\mathfrak{m}F}$ is one that does not generate birefringence. (The dependence of
$k_{\mathfrak{m}F}$ on $\mathfrak{m}^{V}$ is analogous to how the Ricci tensor embeds in the full Riemann
curvature tensor.) This is not an artifact of the leading order approximation either; even with the
full form~\eqref{eq-Lmass}, including terms quadratic in $\mathfrak{m}^{V}$, there is no photon birefringence.
So while we may give some estimates of how tightly the Lorentz-violating contributions to the vector meson
mass matrices may be constrained by purely electromagnetic measurements, the very best electromagnetic bounds
will play no role in the analysis.

According to the hypotheses of VMD, the fully dressed photon state $|\gamma\rangle$ may be expressed as
a superposition
\begin{equation}
\label{eq-gamma-dressed}
|\gamma\rangle\approx\sqrt{Z_{3}}|\gamma_{b}\rangle+\sqrt{\alpha}|h\rangle,
\end{equation}
where $|\gamma_{b}\rangle$ is the bare photon and $|h\rangle$ describes the hadronic contribution.
The factor $\sqrt{Z_{3}}$ is present to assure that the state is properly normalized when the hadronic part
is added, and the explicit $\sqrt{\alpha}$ dependence on the fine structure constant is conventionally included
to indicate that the mixing term is $\mathcal{O}(e)$.
The mixing may be taken to be induced by an photon-meson interaction Lagrange
density~\cite{ref-gell-man2,ref-kroll}
\begin{equation}
\label{eq-Lmix}
\mathcal{L}^{V}_{\mathrm{mix}}=\sum_{V}\frac{em_{V}^{2}}{\tilde{f}_{V}}A^{\mu}V_{\mu},
\end{equation}
in terms of the photon and meson fields $A$ and $V$. $\tilde{f}_{V}$ is a nondimensionalized 
form of the meson decay constant, renormalized to zero momentum,
\begin{equation}
\tilde{f}_{V}^{2}=f_{V}^{2}\left[1-\frac{\Pi_{V}(0)}{m_{V}^{2}}\right],
\end{equation}
where $\Pi_{V}(0)$ is the vacuum polarization contribution to the meson propagator and $f_{V}$
is the true decay constant, evaluated at the meson pole mass. This may be determined from, for example,
the electromagnetic decay rate $\Gamma_{V\rightarrow e^{-}e^{+}}$, via
\begin{equation}
\label{eq-f-correction}
f_{V}^{2}=\frac{4\pi\alpha^{3}}{3}\frac{m_{V}}{\Gamma_{V\rightarrow e^{-}e^{+}}}.
\end{equation}
The correction term in~\eqref{eq-f-correction} is generally relatively small, at the less than 10 percent
level for the $\rho^{0}$~\cite{ref-gounaris}.
The $\rho^{0}$ couples significantly more strongly to the photon than the heavier vector mesons, and for
this field we shall use the conservative estimate $\tilde{f}_{\rho}\lesssim 2.5$.

The effective interaction~\eqref{eq-Lmix} gives rise to a hadronic admixture in a virtual photon with
spacelike four-momentum $-Q^{2}$,
\begin{equation}
\label{eq-lin-comb}
\sqrt{\alpha}|h\rangle=\sum_{V}\frac{e}{f_{V}}\frac{m_{V}^{2}}{Q^{2}+m_{V}^{2}}|V\rangle.
\end{equation}
This result arises straightforwardly at first order in perturbation theory. However, it should be emphasized that
the Lagrange density $\mathcal{L}^{V}_{\mathrm{mix}}$ is not, strictly speaking,
applicable beyond leading order, since it is not
gauge invariant beyond this order. It describes an effective interaction between the electromagnetic field
and a current that is expressed entirely in terms of the vector meson fields.
There are, however, fully gauge-invariant ways of formulating the
photon-meson mixing Lagrange density that agree with~\eqref{eq-Lmix} at their lowest orders~\cite{ref-kroll},
so there is no problem with using~\eqref{eq-Lmix} to derive~\eqref{eq-gamma-dressed}.
The mixing~\eqref{eq-gamma-dressed} is responsible for a number of important physical effects beyond the
modified photon propagation phenomena that are the main focus of this paper. Of particular significance
is the way that the hadronic admixture $\sqrt{\alpha}|h\rangle$ contributes to $\rho^{0}$ photoproduction,
$\gamma+N\rightarrow\rho^{0}+N$. The virtual hadronic component of the dressed photon interacts with the target
nucleon, producing various particles, particularly $\pi^{+}\pi^{-}$ pairs. However, by virtue of the optical
theorem, absorption of the $\rho^{0}$ must be accompanied by at least as much diffractive scattering into the shadow
of the nucleon, which produces an outgoing on-shell $\rho^{0}$

There is also a natural Lorentz-violating generalization of $\mathcal{L}^{V}_{\mathrm{mix}}$. Rather than
contracting the vector indices of the photon field and meson field, a new two-index tensor may intervene,
with the resulting action taking the form
\begin{equation}
\mathcal{L}^{V}_{\mathrm{mix,\,LV}}=\sum_{V}\frac{em_{V}^{2}}{\tilde{f}_{V}}k_{AV}^{\mu\nu}A_{\mu}V_{\nu}.
\end{equation}
This is normalized so that $k_{AV}$ is dimensionless.  $k_{AV}$ is traceless, and in order that this may be
further correctable to preserve gauge invariance, it should be symmetric in its two indices. This term also
can contribute to the effective photon $k_{F}$ term.

\section{Experimental Constraints}
\label{sec-exp}

In section~\ref{sec-Lag}, four different EFT coefficients for the SME generalization of a VMD field theory
have been identified:  $k_{G^{V}}$ and $k_{VG}$, which affect the kinetic term for free propagation of the
vector mesons; $\mathfrak{m}^{V}$, a Lorentz-violating modification of the meson mass term; and $k_{AV}$,
which modifies the photon-meson mixing directly.

While $k_{\mathfrak{m}F}$ depends on a cutoff scale $\Lambda$ and is thus imperfectly controlled, the
other three types of meson-sector coefficients give rise to quantitatively well understood contributions
to the effective photon coefficients $k_{F}$ and $k_{AF}$. These are simply weighted sums of the
corresponding coefficients for the mesons. For the mixing of on-shell photons, these effective coefficients
are simply
\begin{eqnarray}
k_{G^{V}F}^{\mu\nu\sigma\tau} & = & \sum_{V}\frac{4\pi\alpha}{f_{V}^{2}}k_{G^{V}}^{\mu\nu\sigma\tau} \\
& \approx & 0.015\,k_{G^{\rho}}^{\mu\nu\sigma\tau}+0.0001\,k_{G^{\omega}}^{\mu\nu\sigma\tau}
+0.0008\,k_{G^{\phi}}^{\mu\nu\sigma\tau}
\label{eq-kF-combination}
\end{eqnarray}
and
\begin{equation}
k_{VGAF}^{\mu}\approx 0.015\,k_{\rho G}^{\mu}+0.0001\,k_{\omega G}^{\mu}+0.0008\,k_{\phi G}^{\mu},
\label{eq-kAF-combination}
\end{equation}
where the differences between $f_{V}$ and $\tilde{f}_{V}$ are not relevant at this level of precision.

At a strictly phenomenalistic level, coefficient tensors like $k_{G^{\rho}}$ and $k_{\rho G}$ may simply be
seem as elements of the meson-sector SME. However, more fundamentally, they must arise out of the
Lorentz and CPT violation coefficients for the underlying fundamental fields. Interestingly, there are no
quark-sector coefficients that can contribute to the birefringent part of a $k_{G^{V}}$; the portions of
those tensors with the Weyl-tensor-like symmetry must be linear combinations of vector boson
coefficients---meaning $k_{AF}$ itself, as well as its gluon and electroweak analogues, with the gluon
contribution having the largest coefficient in each linear combination. This could be made more precise
with full $\chi PT$ models of Lorentz violation in the isovector and isoscalar vector meson sectors.

The dimension-three and dimension-four coefficients that generate birefringence have been constrained most
tightly by looking at the polarization patterns of radiation that has traversed cosmological distances---from
quasar jets~\cite{ref-carroll1,ref-mewes8,ref-gerasimov}, gamma-ray bursts~\cite{ref-mewes7},
and (most distant of all) the cosmic microwave
background (CMB)~\cite{ref-mewes5,ref-contreras,ref-minami,ref-nilsson2,ref-caloni}. The bounds
on the birefringent $k_{F}$ parameters are all at the $10^{-35}$ level or better, while for the
coefficients associated with the directions to particular sources of polarized $\gamma$-rays, there
are bounds at the better than $10^{-38}$ level. For the $\gamma$-ray bursts, the higher energies more
than compensate for the poorer resolution of the photons' polarization directions. If there is no
fundamental $k_{F}$ parameter in the photon sector, then all these bounds may instead be interpreted
as constraints on the linear combination~\eqref{eq-kF-combination}. The numerical reach for each
coefficient in the $\rho^{0}$ sector is approximately two orders of magnitude looser than
the corresponding bounds for pure photons (meaning at $10^{-33}$--$10^{-36}$ level),
three orders of magnitude looser for the $\phi$ sector ($10^{-32}$--$10^{-35}$),
and four orders of magnitude looser for the $\omega$ ($10^{-31}$--$10^{-34}$).

Bounds for the $k_{AF}$ are similarly stringent. Polarization correlations in the CMB provide
an extremely powerful tool for constraining these coefficients, and all four components of the
electromagnetic $k_{AF}$ are bounded to better than $10^{-45}$ GeV. Bounds much tighter
than this are probably impossible, since the birefringence caused by $k_{AF}$ is independent of
photon energy. The existing bounds are derived from precise measurements made on photons that have
traveled all the way from the last scattering surface---the longest possible line of sight in the universe.
As with the dimension-four terms, the strength of the inferred bounds on the hadronic coefficients are
weaker than the photon bounds by amounts dictated by the vector mesons' weights in the linear
combination~\eqref{eq-kAF-combination}. This means bounds of at least $10^{-43}$ GeV for each
of the four components on $k_{G\rho}$, at least $10^{-42}$ GeV for $k_{G\phi}$, and
$10^{-41}$ GeV for $k_{G\omega}$.

The remaining coefficients all contribute to the electromagnetic sector with the same kind of
Lorentz structure---that which is seen in~\eqref{eq-kmF}. Combining all these terms
into a single effective contribution to the birefringence-free part of the effective $k_{F}$
gives
\begin{equation}
\label{eq-kVF}
k_{VF}^{\mu\nu\sigma\tau}=\frac{1}{2}\left(\tilde{k}^{\mu\sigma}g^{\nu\tau}
-\tilde{k}^{\nu\sigma}g^{\mu\tau}
+\tilde{k}^{\nu\tau}g^{\mu\sigma}
-\tilde{k}^{\mu\tau}g^{\nu\sigma}\right),
\end{equation}
where the Ricci tensor analogue $\tilde{k}^{\mu\nu}=k_{VF\alpha}\,^{\mu\alpha\nu}$ part of the
full effective $k_{F}$ receives contributions
\begin{equation}
\label{eq-k-allF}
\tilde{k}^{\mu\nu}=\sum_{V}\left[\frac{m^{2}_{V}}{2\Lambda_{V}^{4}}\left(\mathfrak{m}^{V}\right)^{\mu\nu}
+\frac{4\pi\alpha}{f_{V}^{2}}\left(k_{G^{V}}\right)_{\alpha}\,^{\mu\alpha\nu}
+\frac{8\pi\alpha}{f_{V}^{2}}k_{AV}^{\mu\nu}\right].
\end{equation}

The best bounds on these nine coefficients, at the $10^{-18}$--$10^{-22}$ levels come from a number of
sources. These types of bounds are always implicitly bounds on differences between SME coefficients in
different sectors; the background $k_{F\alpha}\,^{\mu\alpha\nu}$ describes the natural Cartesian coordinate
system for describing electromagnetism, but with Lorentz violation these electromagnetic coordinates may be
skewed relative to the coordinate systems that are natural for other sectors of the standard model.
There are Michelson-Morely interferometry results, either inferred from LIGO
data~\cite{ref-melissinos} or taken from dedicated resonator experiments~\cite{ref-nagel}.
However, these terrestrial experiments are not as good at bounding coefficients
governing boost invariance violation as they are at constraining anisotropic photon behavior.
High-energy astrophysical observations made on massive quanta boosted to nearly the speed of light
are not subject to this restriction, and so some of the best bounds come from ultra-high-energy cosmic
ray data~\cite{ref-klinkhamer4,ref-duenkel}.

The reductions in sensitivities for the $k_{G^{V}}$ and $k_{G^{V}}$ terms in~\eqref{eq-k-allF}
relative to the photon $k_{F}$ terms are of the same size as for the birefringent terms already discussed:
two order of magnitude of sensitivity lost for $\rho^{0}$ terms, three for $\phi$, and four for $\omega$.
The situation is trickier for the $\mathfrak{m}_{V}$ mass-like parameters, since by introducing them
in~\eqref{eq-kmF} we have moved another layer deeper into EFT and consequently introduced a new
cutoff scale for each species, $\Lambda_{V}$. Realistically, $\Lambda_{V}$ should depend on
$f_{V}$, but it is not clear, with the purely phenomenalistic Lagrange density~\eqref{eq-kmF}, what form
this dependence should take; there is no interesting Lorentz-invariant analogue to $\mathcal{L}_{\mathrm{mass}}$
to compare to. So we shall merely state that bounds on the $\mathfrak{m}^{V}$ components are at the
$\left(10^{-14}\right.$--$\left.10^{-18}\right)(\Lambda_{V}/10\,\mathrm{GeV})^{4}$ levels; these
are still quite stringent bounds even if $\Lambda_{V}\sim100$ GeV.

\section{Conclusion}
\label{sec-concl}

VMD is an extremely important physical phenomenon in photon-hadron interactions. Because it
deals with connections between the photon sector and vector meson sectors of the standard model, it
provides an avenue for constraining exotic forms of new physics for vector hadrons via observations
of photon interactions. Quantitatively, the admixture of meson states into the dressed photon
is small, topping out at approximately 0.015 for $\rho^{0}$ isovector mesons mixed into
on-shell photons. However, since many pure photon tests of Lorentz and CPT symmetry are extremely
strong, the mixing can still provide excellent bounds on effective coefficients for the
mesons themselves.

The bound discussed here are on linear combinations of $\gamma$, $\rho^{0}$, $\phi$, and $\omega$
coefficients. The relative sizes of the contributions are dictated by how strongly each of the
meson mixes with the photon. It would be possible, in principle, to disentangle the contributions
of each individual field by making observations with off-shell photons with different virtual momenta
$-Q^{2}$, according to~\eqref{eq-lin-comb}. However, in practice, the precision of scattering
experiments with spacelike virtual photons is extremely poor compared with the results of photon
birefringence, Michelson-Morley, and cosmic ray tests of Lorentz invariance involving real photons.

VDM provides but one lens for understanding the behavior of the composite hadrons that are the
observable quanta of the strong interaction at low energies. It is worthwhile to understand how the EFTs
developed to study difficult strong interaction regimes may each be modified to accommodate
possible Lorentz and CPT violations. The impact of symmetry-breaking terms on $\chi$PT has already been
explored, albeit incompletely. Incorporation of heavy quark effective theory (HQET) would be a
next natural step; with HQET it will be possible to make quantitative connections between the
results of Lorentz tests on heavy hadrons and the SME coefficients for charm and bottom quarks. A full
understanding of the effective theory of Lorentz and CPT violation for spin-$\frac{3}{2}$ Rarita-Schwinger
particles would also be valuable, since $\Delta$-resonance states play important roles in
ultra-high-energy cosmic ray interactions. Another EFT important for the analysis of strong
interaction physics is soft collinear effective theory (SCET). Although SCET is primarily useful for
the analysis of multiple products emerging from high-energy collisions---not an area where
Lorentz and CPT tests are likely to be particularly precise---it may also be interesting to
examine how SCET might accommodate such symmetry violations. Moreover, for all of these
EFTs, a full understanding of quantum corrections to would also be valuable. There may be
quite a bit of interesting SME physics to be extracted from modifications of existing EFT models.

\end{document}